\documentclass{article}
\usepackage{cite}
\usepackage{spconf,amsmath,graphicx}
\usepackage{algorithmic}
\usepackage{graphicx}
\usepackage{caption}
\usepackage{subfigure}
\usepackage{textcomp}
\usepackage{subfigure}
\usepackage{bm}

\title{Low-Rank Subspace Representation from Optimal Coded-Aperture  for Unsupervised Classification of Hypersepctral Imagery}
\name{Jianchen Zhu$^{1,3}$, Tong Zhang$^{1,2}$, Shengjie Zhao\sthanks{Corresponding author: Shengjie Zhao}$^{1,2,3}$}
\address{$^1$Key Laboratory of Embedded System and Service Computing, Ministry of Education \\ $^2$School of Software Engineering, $^3$School of Electronic and Information Engineering  \\ Tongji University, Shanghai, 201804, CHN \\  Email: zhujianchen@tongji.edu.cn, shengjiezhao@tongji.edu.cn.}
\begin{document}
%
\maketitle
\begin{abstract}
This paper aims at developing a clustering approach with spectral images directly from the compressive measurements of coded aperture snapshot spectral imager (CASSI). Assuming that compressed measurements often lie approximately in low dimensional subspaces corresponding to multiple classes, state of the art methods generally obtains optimal solution for each step separately but cannot guarantee that it will achieve the globally optimal clustering results. In this paper, a low-rank subspace representation (LRSR) algorithm is proposed to perform clustering on the compressed measurements. In addition, a subspace structured norm is added into the objective of low-rank representation problem exploiting the fact that each point in a union of subspaces can be expressed as a sparse linear combination of all other points and that the matrix of the points within each subspace is low rank. Simulation with real dataset illustrates the accuracy of the proposed spectral image clustering approach.
\end{abstract}
\section{Introduction}
\label{sec:intro}
Spectral imaging (SI) techniques combine the 2D imaging and spectroscopy to sense spatial information across a multitude of wavelengths, These data sets can be viewed as three-dimensional (3D) images with two spatial and one spectral dimension. Spectral images have been widely used in remote sensing applications. Traditionally SI techniques require scanning the scene per spatial line or tuning a set of band-pass filters for each required spectral band, which leads to increasing of acquisition time \cite{BoremanClassification,Rueda2016Compressive}. Spectral data cubes are a valuable tool for monitoring the Earth's surface since the different objects in the scene reflect, scatter, absorb, and emit electromagnetic energy in distinctive patterns related to their molecular composition. A commonly used technique in these applications is clustering. Spectral image clustering can be seen as the process of segmenting pixels into corresponding sets which satisfy the requirement that differences between sets are much greater than the differences within sets.   
  
Due to the mixed nature of spectral data, i.e., each pixel contains several materials, the underlying data structure often includes multiple subspaces. Subspace clustering theory can be used to model related problems including the spectral image classification. Different methods for subspace clustering have been developed over the past decade, such as iterative methods \cite{Rahmani2015Innovation,Fan2015Global}, algebraic methods \cite{YiEstimation}, statistical methods \cite{XiStructured}, and spectral clustering based methods \cite{elhamifar2013sparse,soltanolkotabi2014robust}. Among them, low-rank representations (LRR) based methods find a lowest rank representation for subspace clustering, which in turn, results in the global structure of the data as well as be robust to noise \cite{liu2013robust,sumarsono2015low-rank}. However, direct application of the LRR algorithm to spectral images is a challenging task because of the high-dimensional spectral data sets, which require huge computational resources and storage capacities. Therefore, to mitigate these problems, it is necessary to reduce the dimensionality of spectral images.        

Compressive spectral imaging systems require fewer  measuments than those with traditional spectral imaging sensors. Our work aims at the clustring of the data acquired by a compressive imager known as the spatial-spectral coded compressed spectral imager (3D-CASSI) system. The 3D-CASSI system first encodes spatial and spectral information of a scene using a 3D coded aperture and then the coded information is integrated along the spectral dimension. The 3D-CASSI system is different from the system in \cite{martin2016hyperspectral,cao2016computational,arguello2014colored,Yuehao2011Development,Arguello2013Higher} because each spatial position of the acquired measurements contains the compressed information of a single coded spectral signature\cite{Xun2016Computational}.

Assuming that the compressed measurements lie in the union of multiple low-dimensional subspaces, this paper focuses on the unsupervised classification of every spectral pixel of the scene into one of the known classes from the given set of 3D-CASSI compressive measurements，without first reconstructing the full 3-D full datacube. The proposed approach is based on the LRSR model where each spectral signature from its own subspace can be represented by other pixels in the same subspace. Further, similar materials are represented as the neighboring pixels in a spectral image, which can help to extract more information from the data and reduce the representation error by imposing a low-rank constraint to the spare matrix \cite{zhu2019low-rank}. 

The main contributions of this paper are twofold. First, the coded apertures used in the 3D-CASSI are realized by a greedy pursuit (GP) algorithm such that optimal compressed measurements are acquired, allowing the performance of spectral image clustering to be improved on comparison with the clustering performance obtained when traditional randomly coded pastures are used \cite{wagadarikar2008single,Arguello2015Restricted}. Second, a novel low-rank representation based subspace clustering algorithm is proposed to perform the spectral pixel clustering directly in the compressed measurements.

\section{PROBLEM FORMULATION}
\label{sec:format}
\begin{figure*}
	\centering
	\includegraphics[width=14cm, height=5cm]{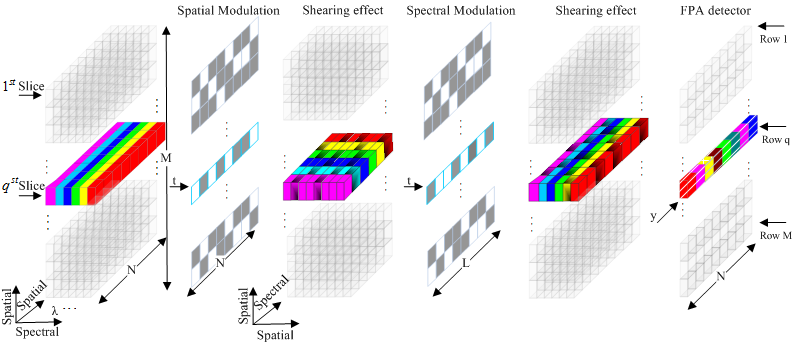}
	\caption{Illustration of the spatial-spectral optical flow in the 3D-CASSI. The $ q^{th} $ slice of the datacube $ F $ with $ L=6 $ spectral components is coded by a row of the coded aperture $ t $ and sheared by the dispersive element. The detector acquires the intensity $ y $ by integrating the coded light.}
	\label{fig:figure2}
\end{figure*}

In the 3D-CASSI system, as shown in Fig. 1, the voxels of the spectral scene is first modulated by a coded aperture. The coded spectral pixels are then integrated in the focal plane array detector (FPA) detector along the spectral axis. Let $ T_{m,n,k}^{s} $ be the time-varying tridimensional coded aperture in its discrete form and $ F_{m,n,k} $ be the spatial-spectral density source, where $ m $, $ n $ index the spatial coordinates, $ k $ the spectral component, and $ s $ the temporal component. The $ s^{th} $ discrete output on the FPA can be expressed as 
\begin{equation}\label{key}
{Y_{m,n}^{s}} = \sum\limits_{k = 0}^{L - 1} {T_{m,n,k}^{s}}{{F_{m,n,k}}}+w_{m,n}，
\end{equation}
where $ {Y_{m,n}^{s}} $ denotes the attained measurement at the $ (m,n)^{th} $ position on the detector at a specific snapshot $ s $ whose dimensions are $ M\times N $ and $ w_{m,n} $ is the white noise of the sensing system. Equation (1) can then be rewritten in a matrix form  
\begin{equation}\label{key}
y^{s}=H^{s}f+e,
\end{equation}
where $ y^{s}\in R^{MN} $ and $ f\in R^{MNL} $ are the vectorized representation of $ {Y_{m,n}^{s}} $ and $ {F_{m,n,k}^{s}} $, respectively, and $ H $ is the measurement matrix of the CASSI system, which is determined by the coded aperture $ {T_{m,n,k}^{s}} $. The ensemble of $ S $ measurements can be expressed as 
\begin{equation}\label{key}
y^{S}=H^{S}f+e,
\end{equation}
where $ y^{S}=[(y^{0})^{T},...,(y^{S-1})^{T}] $ and $ H $ is the concatenation of matrices $ H^{s} $, $ s=0,...,S-1 $. 

Alternatively, considering that preserving the structure of the underlying high dimensional data can further improve the subspaces clustering results directly on the compressed domain, the matrix of $ S $ coding pattern is defined as $ H=[H^{0},H^{1},...,H^{S-1} ]^{T}\in R^{S\times L} $ and $ f=[f_{0}^{T},...,f_{L-1}^{T}]^{T} $ is a $ L\times MN $ matrix whose columns are the spectral signatures $ f_{j} $ of the data cube. The ensemble of $ S $ measurements can be expressed as $ y^{S}=[(y^{0})^{T},...,(y^{S-1})^{T}]^{T} $ where $ y^{S} $ is a $ S\times MN  $ matrix. Notice that in the matrix $ y^{S} $ each column value and each row value correspond to a compressed spectral signature and the compressed information (spectral response) of each pixels obtained at $ s^{th} $ snapshot, respectively. Then, the matrix $ y $ has convenient representation for SSC due to its structure, which makes easy to discriminate among compressed measurements.

\section{Greedy Pursuit Algorithm for Coding Pattern Optimization}
\label{sec:pagestyle}
we first utilize a smooth function of wavelength to obtain the information from the given sets of neighboring spectral bands of interest, which leads to the preservation of the original signal structure \cite{Liu2014Adaptive}. Let $ (\left \{ \lambda _{1}^{S},\lambda _{2}^{S} \right \})=(\left \{ \lambda _{1}^{0},,\lambda _{2}^{0} \right \},...,\left \{ \lambda _{1}^{S-1},,\lambda _{2}^{S-1} \right \}) $ be the set selected to the matrix $ (H^{S})_{k}=\delta _{\left \lfloor \lambda _{1}^{S} \right \rfloor}\delta _{\left \lfloor \lambda _{2}^{S} \right \rfloor}h_{k}^{S} $, then the optimization problem can be expressed as  
\begin{equation}\label{key}
\begin{split}
&\underset{H,\lambda _{1}^{S},\lambda _{2}^{S},h^{s}}{min}f(H)=\left \| (H_{k}^{T})H_{k^{'}} \right \|_{F}^{2}+\left \| H^{s}(H^{s^{'}}) \right \|_{F}^{2} \\
&~~~s.t.~H\in \mathcal{C}_{L,S}, ~(H^{s})_{k}=\delta _{\left \lfloor \lambda _{1}^{s} \right \rfloor}\delta _{\left \lfloor \lambda _{2}^{s} \right \rfloor}h_{k}^{s} \\
&~~~~~~~\Delta (\lambda _{2}^{s}-\lambda _{1}^{s})=\Lambda-1, ~\det(H)\neq 0\\
\end{split}
\end{equation}
for $ k=0,...,L-1 $ and $ s=0,...,S-1 $, where $ (H_{k}^{T})H_{k^{'}} $ and $  H^{s}(H^{s^{'}})^{T} $ respectively collect all the entries outside the diagonal of $ H^{T}H $ and $ HH^{T} $, and $ \Lambda $ denotes the coding pattern bandwidth.

The problem (4) can be solved by applying the greedy pursuit (GP) as shown in Algorithm 1 because it can reduce computational complexity and improves computational efficiency. The interest of using GP is to obtain the optimal projections to solve the spectral image clustering problem. Notice in Fig. 2(a) that the block-unblock entries for the optimal coding pattern present a uniform spectral distribution providing a better sampling. Notice in Fig. 2(b) that the random coding pattern results in oversampling or unsampling of part of all spectral bands. 

\begin{figure}[htb]
	\centering
	
	\subfigure[]{
		\begin{minipage}[t]{0.46\linewidth}
			\centering
			\includegraphics[width=4.3cm, height=3.6cm]{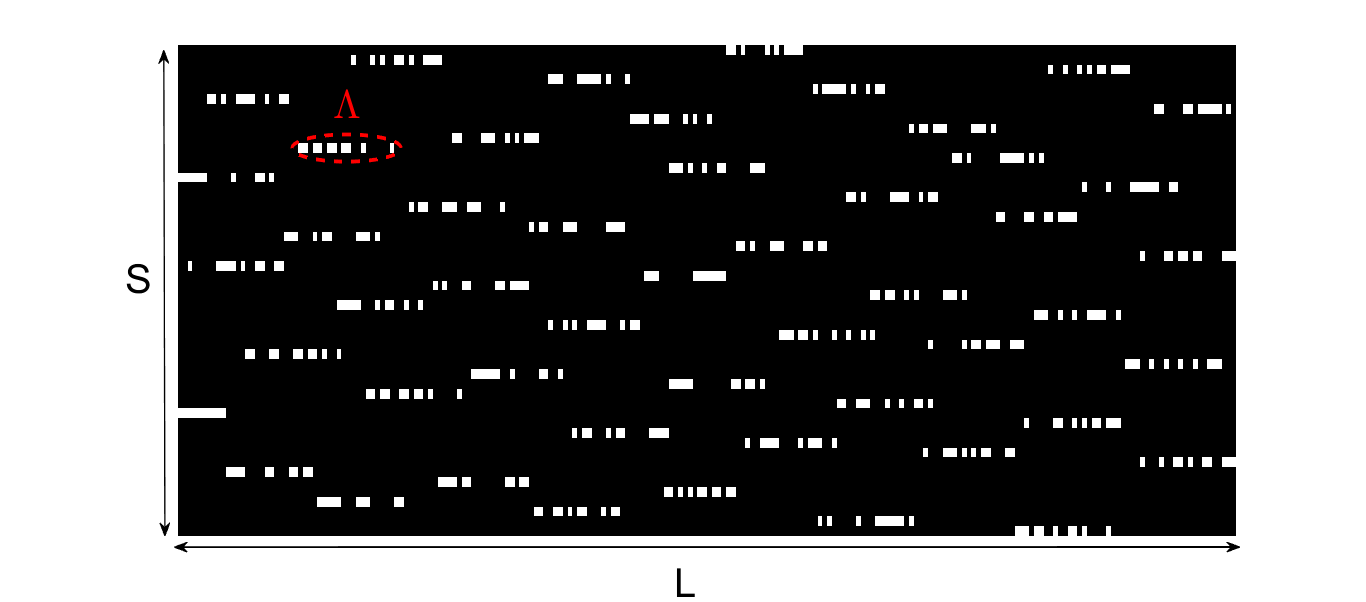}
		\end{minipage}%
	}%
	\subfigure[]{
		\begin{minipage}[t]{0.47\linewidth}
			\centering
			\includegraphics[width=4.3cm, height=3.6cm]{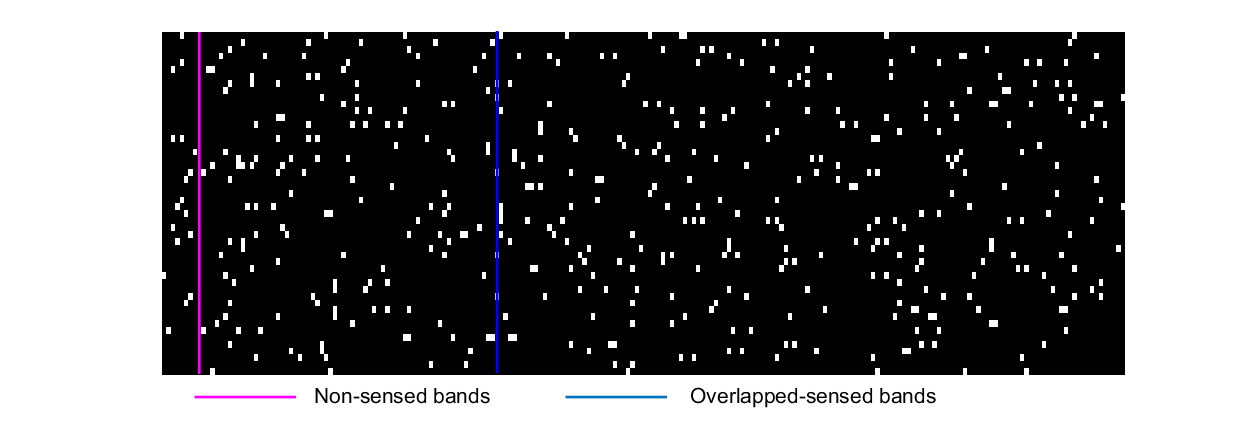}
		\end{minipage}%
	}%
	
	\centering
	\caption{Examples of (a) the optimal clustering coding pattern and (b) the random coding pattern. }
\end{figure}

\begin{table}[htb]
	\centering
	\label{}
	\begin{tabular}{p{2.9in}}
		\hline
		$\bm{ Algorithm }$ $ \bm{ 1 } $$ : $ Generate the optimal coding pattern  \\ \hline
		$\bm{ Input }$$ : $ $ L $, $ S $, $ \Lambda  $                             \\
		$\bm{ Output }$$ : $ $ H\in \left \{ 0,1 \right \}^{S\times L} $                            \\
		1: Initialize: $ (H^{0})_{k}\leftarrow \delta _{\left \lfloor \lambda _{1}/k \right \rfloor}\delta _{\left \lfloor k/\lambda _{2} \right \rfloor}h_{k}^{0}\sim Be(\frac{1}{2}) $                                 \\
		2: $\bm{ for }$ $ s_{it}\leftarrow 1 $ to $ S-1 $ $\bm{ do }$                                               \\
		3:~~~~~~$\bm{ for }$ $ k^{'}\leftarrow 0 $ to $ (L-\Lambda ) $ $\bm{ do }$                                              \\
		4:~~~~~~~~~~$ u_{k^{'}}\leftarrow \sum_{s^{'}=0}^{s_{it}}\sum_{k=k^{'}}^{k^{'}+\Lambda -1}(H^{s^{'}})_{k} $                                              \\
		5:~~~~~~$\bm{ end }$ $\bm{ for }$                                              \\
		6:~~~~~$ \lambda _{1}^{s_{it}},~\lambda _{2}^{s_{it}}\sim U_{ran}[argmin_{k^{'}}u_{k^{'}}]   $                        \\
		7:~~~~~~$ k^{'}k^{'}=0 $                      \\
		8:~~~~~~$\bm{ for }$ $ k^{'} \leftarrow \lambda _{1}^{s_{it}}~to~\lambda _{2}^{s_{it}} $                                              \\
		9:~~~~~~$ u_{k^{'}k^{'}}\leftarrow \sum_{s^{'}=0}^{s_{it}}\prod_{k=(k^{'}-1)}^{k^{'}}(H^{s^{'}})_{k} $, with $ h_{k^{'}}^{s_{it}}\leftarrow 0 $                                              \\ 
		10:~~~~~$ k^{'}k^{'}=k^{'}k^{'}+1 $                                              \\ 
		11:~~~~~$\bm{ end }$ $\bm{ for }$                                             \\
		12:~~~~~$\bm{ for }$ $ {k^{''}}\leftarrow 0 $ to $ \left \lfloor \frac{1}{2}\Lambda  \right \rfloor  $ $\bm{ do }$                                           \\ 
		13:~~~~~~~~~~$ \Gamma \sim U_{ran}^{'}[argmin_{k^{'}k^{'}}u_{k^{'}k^{'}}] $          \\
		14:~~~~~$ h_{\lambda_{l}^{s_{it}}}^{s_{it}}\sim Be(\frac{1}{2}) $, with $ \lambda _{l}^{s_{it}}\subset (\lambda _{1}^{s_{it}},\lambda _{2}^{s_{it}}) $                                                  \\
		15:~~~~~$\bm{ end }$ $\bm{ for } $                                           \\
		16:~~~~~$ (H^{s_{it}})\leftarrow \delta _{\left \lfloor \lambda_{1}^{s_{it}}/k \right \rfloor}\delta _{\left \lfloor k/\lambda_{2}^{s_{it}} \right \rfloor}h_{k}^{s_{it}} $                                         \\
		17:~$\bm{ end }$ $\bm{ for }$               \\   \hline 
	\end{tabular}
\end{table}

\section{Low-Rank and Structured Sparse Subspace Clustering Algorithm for CSI}
\label{sec:typestyle}
Assuming that compressed spectral pixels of the same land-cover class lie in one independent subspace, subspace clustering based methods can be used to separate them into the same cluster. Besides, considering that a specific land-cover material should be regionally distributed in the image, their representation coefficients should also be very close. In particular, LRSR builds the similarity matrix, which describes the relationships between data points exploiting the fact that each compressed pixel is represented as a linear combination of other points in the same subspace. Once obtained the compressive measurements $ y=Hf $, the LRSR seeks a low-rank representation by solving an optimization problem as follows
\begin{equation}\label{key}
\underset{c,g}{min}\left \| c \right \|_{*}+\frac{\lambda}{2} \left \| g \right \|_{2,1}，~s.t. ~y=yc+g, ~diag(z)=0,\\
\end{equation}
where $ c\in R^{MN\times MN} $ refers to the coefficients matrix and the nuclear norm regularization$ \left \| \cdot  \right \|_{*} $ suggest that a low-rank representation of a data point from the same subspace. The matrix $ g $ denotes the representation error, $ \left \| \cdot  \right \|_{2.1} $-norm guarantees that LRSR can captures the global structure of the points as well as robust to noise, and $ \lambda>0  $ is a tradeoff parameter. The constraint $ diag(c)=0 $ is used to eliminate the trivial ambiguity where a data point is expressed by itself. Next, let $ q=[q_{1},...,q_{\kappa  }]\in \left \{ 0,1 \right \}^{MN\times \kappa } $ be the membership of each data point to each subspace. Assuming that the subspace number is $ \kappa $ and  $ rank(q)=\kappa $, the space of segmentation matrices can thus be expressed as
\begin{equation}\label{key}
q=\left \{q\in \left \{0,1\right \}^{MN\times \kappa}:~q1=1~and~ranl(q)=\kappa\right \},
\end{equation}
where $ q1=1 $ indicates that each data point lies in only one subspace with 1 being the vector of all ones of appropriate dimension. A binary matrix $ \Theta \in R^{MN\times MN} $ is introduced to show whether a pair of data points lies in the same subspace. Notice that the coefficients matrix $ c $  encodes the information for segmenting the data and the binary matrix $ \Theta $ can be expressed as $ \left \| c \right \|_{q}=\left \|\Theta \odot c   \right \|_{1} $, which is called subspace structured norm. Moreover, subspace structured norm $ \left \| c \right \|_{q} $ bridge the gap between the low-rank representation $ c $ and the segmentation matrix $ q $. Then, the problem of finding a low-rank representation coefficient matrix exploiting the segmentation of the data points is formulated as the joint optimization problem
\begin{equation}\label{key}
\begin{split}
&\underset{c,g,z}{min}\left \| z \right \|_{*}+\alpha \left \| \Theta \odot c \right \|_{1}+\lambda \left \| g \right \|_{2,1} \\
&s.t.~y=yc+g, ~diag(z)=0,~c=z, \\
\end{split}
\end{equation}
where $ \alpha $ is a tradeoff parameter. The minimization can be efficiently solved by the linearized alternating direction method (LADM). Specifically, when computing the similarity matrix $ w=\left | z \right |+\left | z \right |^{T}\in R^{MN\times MN}  $, the final clustering results is obtained by applying spectral clustering to it.

\section{SIMULATION RESULTS AND DISCUSSION}

The University of Pavia image, was acquired by the Reflective Optics System Imaging Spectrometer System (ROSIS) sensor over the University of Pavia, Pavia, Italy. The size of the image is $ 140\times 80 $, with $ 115 $ bands containing eight main land-cover classes:  asphalt, meadows, trees, metal sheet, bare soil, bitumen, bricks, and shadows. The spectral curves of the eight land-cover classes are shown in Fig. 3. The false-color image and the ground truth are also provided. 

\begin{figure}[htbp]
	\centering
	\subfigure[]{
		\begin{minipage}[t]{0.25\linewidth}
			\centering
			\includegraphics[width=2.3cm, height=2.7cm]{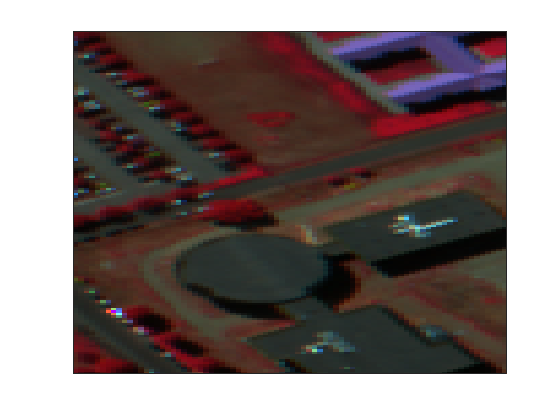}
		\end{minipage}%
	}%
	\subfigure[]{
		\begin{minipage}[t]{0.25\linewidth}
			\centering
			\includegraphics[width=2.3cm, height=2.7cm]{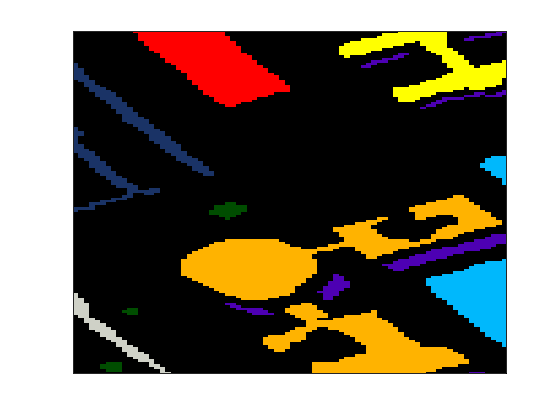}
		\end{minipage}%
	}%
	\subfigure[]{
		\begin{minipage}[t]{0.4\linewidth}
			\centering
			\includegraphics[width=3.5cm, height=2.8cm]{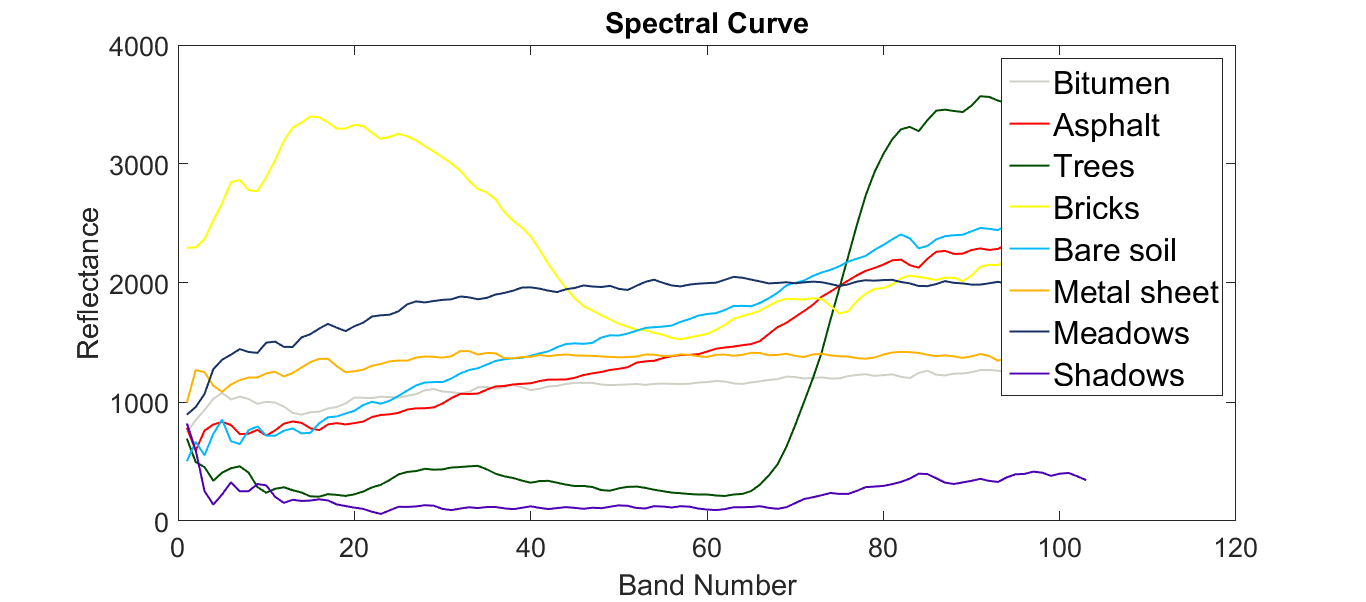}
		\end{minipage}
	}%
	\centering
	\caption{ ROSIS University of Pavia image, (a) False-color image (RGB 102, 56, 31). (b) Ground truth. (c) Spectral curves of the eight land-cover classes.}
\end{figure}

\begin{figure}[htb]
	\centering
	
	\subfigure[]{
		\begin{minipage}[t]{0.3\linewidth}
			\centering
			\includegraphics[width=3cm, height=3cm]{figure/figure25}
		\end{minipage}%
	}%
	\subfigure[]{
		\begin{minipage}[t]{0.3\linewidth}
			\centering
			\includegraphics[width=3cm, height=3cm]{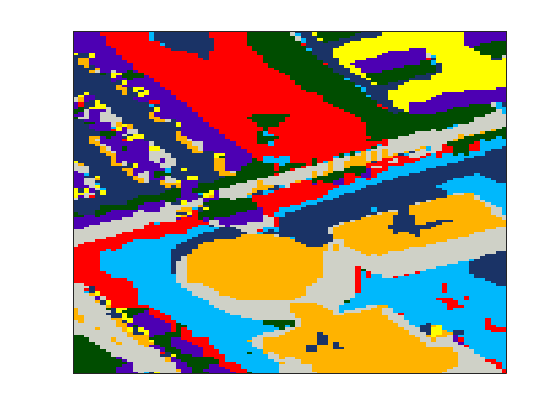}
		\end{minipage}%
	}%
	\subfigure[]{
		\begin{minipage}[t]{0.3\linewidth}
			\centering
			\includegraphics[width=3cm, height=3cm]{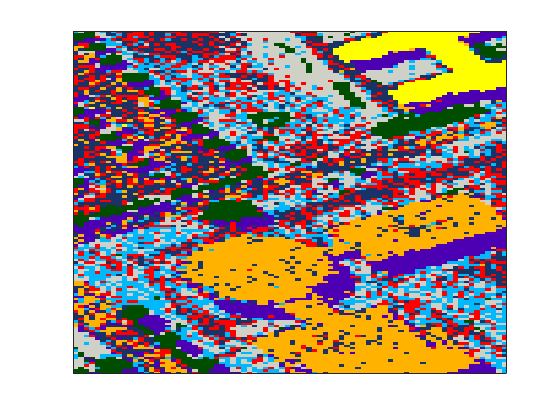}
		\end{minipage}%
	}%

	\subfigure[]{
		\begin{minipage}[t]{0.3\linewidth}
			\centering
			\includegraphics[width=3cm, height=3cm]{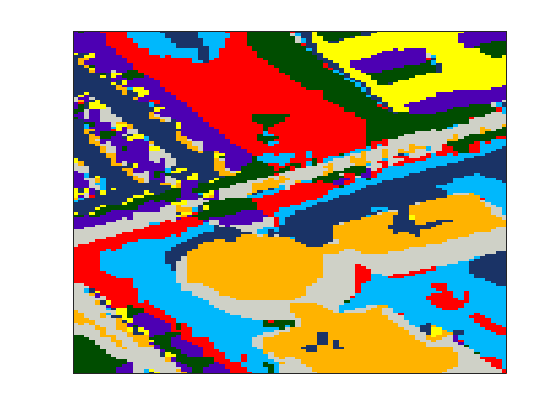}
		\end{minipage}
	}%
	\subfigure[]{
		\begin{minipage}[t]{0.3\linewidth}
			\centering
			\includegraphics[width=3cm, height=3cm]{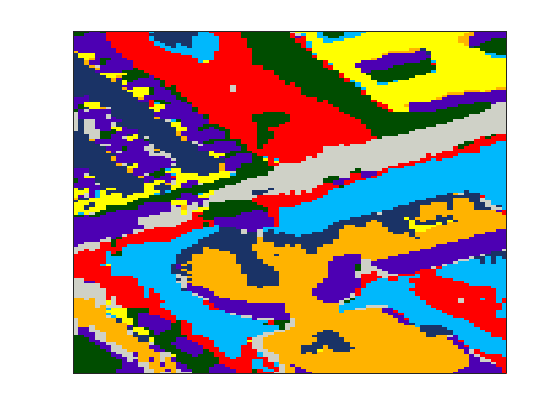}
		\end{minipage}
	}%
	\subfigure{
	\begin{minipage}[t]{0.35\linewidth}
		\centering
		\includegraphics[width=2.7cm, height=2.5cm]{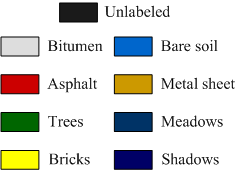}
	\end{minipage}
}%
	
	\centering
	\caption{ Cluster maps of different approaches with the University of Pavia image: (a) Ground truth. (b) Full-data-LRSR-SS, (c) Full-data-LRSR, (d) Optimal-codes-LRSR-SS, (e) Random-codes-LRSR-SS.}
\end{figure}

\begin{table}[htb]
	\centering
	\captionsetup{font=scriptsize}
	\caption{QUANTITATIVE REALUATION OF THE DIFFERENT CLUSTERING APPROACHES FOR THE UNIVERSITY OF PAVIA IMAGE}
	\label{tab:my-table}
	\tabcolsep 0.03in
	\begin{tabular}{ccccc}
		\hline
		Class        & Random & Optimal & Full-LRSR & Full-LRSR-SS \\ \hline
		Asphalt            & $ \underline{71.27} $  & 68.34   & 33.84         & $ \mathbf{80.26} $              \\
		Meadows            & $ \underline{91.17} $  & 100     & 55.02         & $ \mathbf{100} $                \\
		Tree            & $ \underline{90.38} $  & 89.64   & $ \mathbf{100} $  &  $ \underline{90.38} $              \\
		Metal sheets            & 82.90  & $ \mathbf{98.83} $   & $ \underline{91.00} $         & 97.73              \\
		Bare soil            & 46.78  & $ \underline{62.79} $   & 36.26         & $ \mathbf{66.67} $              \\
		Bitumen            & 18.60  & $ \underline{89.57} $   & 0             & $ \mathbf{90.70} $              \\
		Bricks            & $ \mathbf{100} $    & $ \underline{99.68} $   & $ \underline{99.68} $         & $ \underline{99.68} $              \\
		Shadows            & $ \mathbf{99.48} $  & 25.63   & $ \underline{98.45} $         & 24.35              \\ \hline
		OA           & 78.72  & $ \underline{84.85} $   & 71.45         & $ \mathbf{86.85} $              \\
		AA           & 75.09  & $ \underline{79.38} $   & 64.28         & $ \mathbf{81.22} $              \\
		Kappa        & 72.63  & $ \underline{79.52} $   & 62.95         & $ \mathbf{82.50} $              \\ \hline
		Time {[}s{]} & 587.63 & 135.82  & 5214.8        & 14646.9            \\ \hline
	\end{tabular}
\end{table}

The clustering results of different approaches are provided in Fig. 4 and Table 1 (including overall accuracy (OA), average accuracy (AA), Kappa coefficients, and computational time (Time)), respectively. In Table 1, the best result and the second-best result of each row are shown in bold and underlined, respectively. From Fig. 4 and Table 1 it can be seen that the low-rank subspace represtentation plus subspace structured (LRSR-SS) , using the optimal clustering coding patterns, provide pretty similar results to applying clustering driectly on the full 3D datacube. Further, the proposed approach provdes the shortest clustering time.

\section{CONCLUSION}
A compressed spectral image clustering approach has been derived, which bypasses the computational expensive task of applying clustering directly on the full spectral data cube to then apply clustring dircetly from the CASSI measurements. Simulations show the proposed approach achieves relatively well accuracy, but it is up to $ 10 $ times faster than other approaches.

\section{ACKNOWLEDGMENT}
This work was supported in part by the National Natural Science Foundation of China under Grant 61936014, and in part by the National Key Research and Development Project under Grants 2019YFB2102300 and 2019YFB2102301, and was supported by the Fundamental Research Funds for the Central Universities.

\bibliographystyle{IEEEtran}
\bibliography{reference1}

\begin{thebibliography}{10}
\providecommand{\url}[1]{#1}
\csname url@samestyle\endcsname
\providecommand{\newblock}{\relax}
\providecommand{\bibinfo}[2]{#2}
\providecommand{\BIBentrySTDinterwordspacing}{\spaceskip=0pt\relax}
\providecommand{\BIBentryALTinterwordstretchfactor}{4}
\providecommand{\BIBentryALTinterwordspacing}{\spaceskip=\fontdimen2\font plus
\BIBentryALTinterwordstretchfactor\fontdimen3\font minus
  \fontdimen4\font\relax}
\providecommand{\BIBforeignlanguage}[2]{{%
\expandafter\ifx\csname l@#1\endcsname\relax
\typeout{** WARNING: IEEEtran.bst: No hyphenation pattern has been}%
\typeout{** loaded for the language `#1'. Using the pattern for}%
\typeout{** the default language instead.}%
\else
\language=\csname l@#1\endcsname
\fi
#2}}
\providecommand{\BIBdecl}{\relax}
\BIBdecl

\bibitem{BoremanClassification}
Boreman and G.~D., ``Classification of imaging spectrometers for remote sensing
  applications,'' \emph{Optical Engineering}, vol.~44, no.~1, p. 013602.

\bibitem{Rueda2016Compressive}
H.~Rueda, H.~Arguello, and G.~R. Arce, ``Compressive spectral testbed imaging
  system based on thin-film color-patterned filter arrays,'' \emph{Applied
  Optics}, vol.~55, no.~33, p. 9584, 2016.

\bibitem{Rahmani2015Innovation}
M.~Rahmani and G.~Atia, ``Innovation pursuit: A new approach to subspace
  clustering,'' \emph{IEEE Transactions on Signal Processing}, vol.~PP, no.~99,
  pp. 1--1, 2015.

\bibitem{Fan2015Global}
Y.~Fan, R.~He, and B.~G. Hu, ``Global and local consistent multi-view subspace
  clustering,'' in \emph{2015 3rd IAPR Asian Conference on Pattern Recognition
  (ACPR)}, 2015.

\bibitem{YiEstimation}
M.~Yi, A.~Y. Yang, H.~Derksen, and R.~Fossum, ``Estimation of subspace
  arrangements with applications in modeling and segmenting mixed data,''
  \emph{Siam Review}, vol.~50, no.~3, pp. 413--458.

\bibitem{XiStructured}
P.~Xi, F.~Jiashi, X.~Shijie, Y.~Wei-Yun, Z.~J. Tianyi, and Y.~Songfan,
  ``Structured autoencoders for subspace clustering,'' \emph{IEEE Transactions
  on Image Processing}, pp. 1--1.

\bibitem{elhamifar2013sparse}
E.~Elhamifar and R.~Vidal, ``Sparse subspace clustering: Algorithm, theory, and
  applications,'' \emph{IEEE Transactions on Pattern Analysis and Machine
  Intelligence}, vol.~35, no.~11, pp. 2765--2781, 2013.

\bibitem{soltanolkotabi2014robust}
M.~Soltanolkotabi, E.~Elhamifar, and E.~J. Candes, ``Robust subspace
  clustering,'' \emph{Annals of Statistics}, vol.~42, no.~2, pp. 669--699,
  2014.

\bibitem{liu2013robust}
G.~Liu, Z.~Lin, S.~Yan, J.~Sun, Y.~Yu, and Y.~Ma, ``Robust recovery of subspace
  structures by low-rank representation,'' \emph{IEEE Transactions on Pattern
  Analysis and Machine Intelligence}, vol.~35, no.~1, pp. 171--184, 2013.

\bibitem{sumarsono2015low-rank}
A.~Sumarsono and Q.~Du, ``Low-rank subspace representation for estimating the
  number of signal subspaces in hyperspectral imagery,'' \emph{IEEE
  Transactions on Geoscience and Remote Sensing}, vol.~53, no.~11, pp.
  6286--6292, 2015.

\bibitem{martin2016hyperspectral}
G.~Martin and J.~M. Bioucasdias, ``Hyperspectral blind reconstruction from
  random spectral projections,'' \emph{IEEE Journal of Selected Topics in
  Applied Earth Observations and Remote Sensing}, vol.~9, no.~6, pp.
  2390--2399, 2016.

\bibitem{cao2016computational}
X.~Cao, T.~Yue, X.~Lin, S.~Lin, X.~Yuan, Q.~Dai, L.~Carin, and D.~J. Brady,
  ``Computational snapshot multispectral cameras: Toward dynamic capture of the
  spectral world,'' \emph{IEEE Signal Processing Magazine}, vol.~33, no.~5, pp.
  95--108, 2016.

\bibitem{arguello2014colored}
H.~Arguello and G.~R. Arce, ``Colored coded aperture design by concentration of
  measure in compressive spectral imaging,'' \emph{IEEE Transactions on Image
  Processing}, vol.~23, no.~4, pp. 1896--1908, 2014.

\bibitem{Yuehao2011Development}
W.~Yuehao, I.~O. Mirza, G.~R. Arce, and D.~W. Prather, ``Development of a
  digital-micromirror-device-based multishot snapshot spectral imaging
  system,'' \emph{Optics Letters}, vol.~36, no.~14, pp. 2692--4, 2011.

\bibitem{Arguello2013Higher}
H.~Arguello, H.~Rueda, Y.~Wu, D.~W. Prather, and G.~R. Arce, ``Higher-order
  computational model for coded aperture spectral imaging,'' \emph{Applied
  Optics}, vol.~52, no.~10, pp. D12--D21, 2013.

\bibitem{Xun2016Computational}
C.~Xun, Y.~Tao, L.~Xing, S.~Lin, and D.~J. Brady, ``Computational snapshot
  multispectral cameras: Toward dynamic capture of the spectral world,''
  \emph{IEEE Signal Processing Magazine}, vol.~33, no.~5, pp. 95--108, 2016.

\bibitem{zhu2019low-rank}
X.~Zhu, S.~Zhang, Y.~Li, J.~Zhang, L.~Yang, and Y.~Fang, ``Low-rank sparse
  subspace for spectral clustering,'' \emph{IEEE Transactions on Knowledge and
  Data Engineering}, vol.~31, no.~8, pp. 1532--1543, 2019.

\bibitem{wagadarikar2008single}
A.~A. Wagadarikar, R.~John, R.~Willett, and D.~J. Brady, ``Single disperser
  design for coded aperture snapshot spectral imaging,'' \emph{Applied Optics},
  vol.~47, no.~10, 2008.

\bibitem{Arguello2015Restricted}
H.~Arguello and G.~R. Arce, ``Restricted isometry property in coded aperture
  compressive spectral imaging,'' in \emph{IEEE International Conference on
  Image Processing}, 2015.

\bibitem{Liu2014Adaptive}
H.~Liu, S.~Liu, Z.~Zhang, J.~Sun, and J.~Shu, ``Adaptive total variation-based
  spectral deconvolution with the split bregman method.'' \emph{Applied
  Optics}, vol.~53, no.~35, pp. 8240--8, 2014.

\end{thebibliography}

\end{document}